\documentclass[letterpaper, 10 pt, conference]{ieeeconf}
\IEEEoverridecommandlockouts                              

\usepackage{amsmath,amsfonts,amssymb}%
\usepackage{bm}%
\usepackage{graphicx}%
\usepackage{cases}%
\usepackage{cite,url,citesort}
\usepackage{verbatim,latexsym}

\newtheorem{theorem}{Theorem}
\newtheorem{remark}{Remark}

\newtheorem{lemma}{Lemma}

\begin{document}
%
\title{\LARGE \bf
 Performance Analysis and Coherent Guaranteed Cost Control for Uncertain Quantum Systems}

\author{Chengdi Xiang, Ian R. Petersen and Daoyi Dong
\thanks{Chengdi Xiang, Ian R. Petersen and Daoyi Dong are with the School of Engineering and Information Technology, University of New South Wales at the Australian Defence Force Academy, Canberra ACT 2600, Australia. {\tt\small \{elyssaxiang, i.r.petersen, daoyidong\}@gmail.com}}%
}


\maketitle

\begin{abstract}
This paper presents several results on performance analysis for a class of uncertain linear quantum systems subject to either quadratic or non-quadratic perturbations in the system Hamiltonian. Also, coherent guaranteed cost controllers are designed for the uncertain quantum systems to achieve improved control performance. The coherent controller is realized by adding a control Hamiltonian to the quantum system and its performance is demonstrated by an example.
\end{abstract}

\IEEEpeerreviewmaketitle

\section{INTRODUCTION}\label{introduction}
Recent years have seen a rapid development of quantum technology and consequently there has been a considerable amount of research focusing on the area of quantum feedback control systems; e.g., see [1]-[7], [10]-[12]. In this research, robustness plays a vital role; e.g., see \cite{matt2008}, \cite{M2003}, \cite{2M2003}. Several robust control methods that are widely used in classical systems have been adopted in quantum control areas. For example, $H^\infty$ control theory has been used to solve a robust feedback controller synthesis problem for quantum systems \cite{matt2008}. A transfer function approach has also been used to analyse robustness in the feedback control of quantum systems\cite{M2003}, \cite{2M2003}. The small gain theorem has been used to analyse the stability and robustness of quantum feedback networks \cite{Helon2006}. In this paper, we extend some results in classical control on system performance analysis and guaranteed cost control design  to quantum systems.

A majority of existing papers in quantum feedback control only consider the case where the controller is a classical system. That is, the controller may be implemented via analog or digital electronics and  quantum measurements are involved.
 However, some recent results have shown that the controller itself can be a quantum system, which is often referred to as coherent quantum control \cite{matt2008}, \cite{M2003}, \cite{2M2003}, \cite{Helon2006}, \cite{Nurdin2009}. The advantage of using coherent quantum control is its ability to achieve improved system performance, since quantum measurements inherently involve the destruction of quantum information.
Such controllers are often defined by linear quantum stochastic differential equations (QSDEs) and require physical realizability conditions so that they represent physically implementable systems; e.g., \cite{matt2008}, \cite{Nurdin2009}, \cite{Maalouf2009}.
The coherent quantum controller, designed in this paper uses the framework involving triples $(S, L, H)$,  where $S$ is a scattering matrix, $L$ is a vector of coupling operators and $H$ is a Hamiltonian operator \cite{matt2009}. The matrix $S$, together with the vector $L$, specifies the interface between the system and the fields and the parameter $H$ describes the self-energy of the system. To control such a quantum system, we add a controller Hamiltonian to the system. In this approach, we do not need to consider physical realizability conditions for the controller system, since a triple $(S, L, H)$ automatically represents a physically realizable quantum system. This $(S, L, H)$ paramerization for open quantum systems  is used in some recent papers, e.g., see \cite{matt2010}, \cite{matt2008}, \cite{Ian2010} and \cite{J2010}, but few controller design methods have been established based on the $(S, L, H)$ approach.

The paper \cite{matt2010} presents conditions of dissipativity and stability for this class of quantum systems. Then, the paper \cite{Ian2012} built on the result of \cite{matt2010} to provide stability conditions for a class of uncertain quantum systems subject to unknown perturbations. Based on \cite{Ian2010} and \cite{matt2010}, we extend the guaranteed cost control method to the quantum domain and provide a performance guarantee for the given system when the system Hamiltonian is in the form of $H=H_1+H_2$, where $H_1$ is a known nominal Hamiltonian and $H_2$ is a perturbation Hamiltonian. Furthermore, motivated by \cite{Ian1994} and \cite{Chang1972}, we add a quantum controller $H_3$ in the system Hamiltonian of the given system not only to guarantee that the system is stable but also to obtain an adequate level of performance.

We begin in Section \ref{quantum systems} by presenting the general class of uncertain quantum system models under consideration. In particular, we specify the underlying systems as linear quantum systems. In Section \ref{perturbations of the hamiltonian}, we present a  class of quadratic perturbation Hamiltonians and a general class of non-quadratic perturbation Hamiltonians. In Section \ref{performance analysis}, we present the performance analysis problem for the given systems in terms of a strict bounded real condition. In Section \ref{optimal quantum controllers design}, we add a quantum controller to the original system to achieve stability and a guaranteed performance level.
In Section \ref{illustrative example}, we provide an example to illustrate the theory which has been developed in this paper.
Conclusions are presented in Section \ref{conclusion}.




\section{QUANTUM SYSTEMS}\label{quantum systems}
The open quantum systems under consideration are defined by parameters $(S, L, H)$ where the system Hamiltonian is decomposed as $H=H_1+H_2$. Here $H_1$ is a known nominal Hamiltonian and $H_2$ is a perturbation Hamiltonian contained in a specified set of Hamiltonians $\mathcal{W}$; e.g.,  \cite{Ian2012},  \cite{matt2010} and \cite{matt2009}.  We define the corresponding generator operator
\begin{equation}
\mathcal{G}(X)=-i[X,H]+\mathcal{L}(X)
\end{equation}
where $\mathcal{L}(X)=\frac{1}{2}L^\dag[X,L]+\frac{1}{2}[L^\dag,X]L$.
Here, $[X,H]=XH-HX$ describes the commutator between two operators and the notation $^\dag $ refers to the adjoint transpose of a vector of operators. $H_1$ and $H_2$ are two self-adjoint operators on the underlying Hilbert space. By introducing a quantum stochastic differential equation, the Heisenberg evolution $X(t)$ of an operator $X$ is defined by the triple $(S,L,H)$ together with the corresponding generators\cite{matt2010}. The results presented in this paper will build on the following results from \cite{Valery}.
\begin{lemma}\label{1lemma}
\cite{Valery} Consider an open quantum system defined by $(S, L, H)$ and suppose there exist non-negative self-adjoint operators $V$ and $W$ on the underlying Hilbert space such that
\begin{equation}\label{lemma1}
\mathcal {G}(V)+W\leq \lambda
\end{equation}
where $\lambda$ is a real number. Then for any plant state, we have
\begin{equation}
\limsup \limits_{T\to \infty} \frac{1}{T}\int_0^T\langle W(t)\rangle dt\leq\lambda.
\end{equation}
Here $W(t)$ denotes the Heisenberg evolution of the operator $W$ and $\langle\cdot\rangle$ denotes quantum expectation; e.g., see \cite{Valery} and \cite{matt2010}.
\end{lemma}


In this paper, we consider nominal systems corresponding to linear quantum systems. We assume that $H_1$ is in the following form
\begin{equation}\label{17}
H_1=\frac{1}{2}\left[\begin{array}{c c} a^\dag & a^T\end{array}\right]M\left[\begin{array}{l} a \\ a^\#\end{array}\right]
\end{equation}
where $M\in \mathbb{C}^{2n\times2n}$  is a Hermitian matrix and has the following form with $M_1=M^\dag_1$ and $M_2=M^T_2$
\begin{equation}
M=\left[\begin{array}{cc}M_1&M_2\\M_2^\#&M_1^\#\end{array}\right].
\end{equation}
Here $a$ is a vector of annihilation operators on the underlying Hilbert space and $a^\#$ is the corresponding vector of creation operators. In the case of matrices, the notation $ ^\dag$ refers to the complex conjugate transpose of a matrix. In the case of vectors of operators, the notation $ ^\#$ refers to the vector of adjoint operators and in the case of complex matrices, this notation refers to the complex conjugate matrix.
The commutation relations between annihilation and creation operators are described as follows
\begin{equation}\label{commutation1}
\begin{split}
\left[\begin{array}{c}\left[\begin{array}{l}a\\a^\#\end{array}\right] ,\left[\begin{array}{l}a\\a^\#\end{array}\right]^\dag \end{array}\right]=&\left[\begin{array}{l}a\\a^\#\end{array}\right] \left[\begin{array}{l}a\\a^\#\end{array}\right]^\dag\\
&-\left(\begin{array}{c}\left[\begin{array}{l}a\\a^\#\end{array}\right]^\#\left[\begin{array}{l}a\\a^\#\end{array}\right]^T \end{array}\right)^T\\
=&\ J
\end{split}
\end{equation}
where $J=\left[\begin{array}{cr}I&0\\0&-I \end{array} \right]$ \cite{Ian2010}.

The coupling vector $L$ is assumed to be of the form
\begin{equation}\label{19}
L=\left[\begin{array}{c c} N_1&N_2\end{array}\right]\left[\begin{array}{l} a \\ a^\#\end{array}\right]
\end{equation}
where $N_1\in \mathbb{C}^{m\times n}$ and $N_2\in \mathbb{C}^{m\times n}$.
We also write
\begin{equation}
\left[\begin{array}{l}L\\L^\#\end{array}\right]=N\left[\begin{array}{l}a\\a^\#\end{array}\right]=\left[\begin{array}{c c} N_1&N_2\\N^\#_2&N^\#_1\end{array}\right]\left[\begin{array}{l} a \\ a^\#\end{array}\right].
\end{equation}

We consider a self-adjoint "Lyapunov" operators $V$ of the form
\begin{equation}\label{v}
V=\left[\begin{array}{c c} a^\dag & a^T\end{array}\right]P\left[\begin{array}{l} a \\ a^\#\end{array}\right]
\end{equation}
where $P\in \mathbb{C}^{2n\times2n}$  is a positive definite Hermitian matrix of the form
\begin{equation}\label{18}
P=\left[\begin{array}{cc}P_1&P_2\\P_2^\#&P_1^\#\end{array}\right].
\end{equation}

We then consider the set of non-negative self-adjoint operators $\mathcal{P}$ defined as
\begin{equation}
\mathcal{P}=\left\{\begin{array}{c} V\ \textnormal{ of the form} \ (\ref{v})\ \textnormal{such that} \ P>0\ \textnormal {is a}\\ \textnormal{Hermitian matrix of the form}\ (\ref{18})\end{array} \right\}.
\end{equation}


\section{PERTURBATIONS OF THE HAMILTONIAN}\label{perturbations of the hamiltonian}
\subsection{Quadratic Hamiltonian Perturbations}
For the set of non-negative self-adjoint operators $\mathcal{P}$ and given real parameters $\gamma>0, \delta\geq0$, a particular set of perturbation Hamiltonians $\mathcal{W}_1$ is defined in terms of commutator decomposition
\begin {equation}\label{decomposition1}
[V,H_2]=[V, z^\dag]w-w^\dag[z,V]
\end{equation}
for $V\in \mathcal{P}$, where $w$ and $z$ are given vectors of operators.
$\mathcal{W}_1$ is then defined in terms of sector bound condition:
\begin{equation}\label{sector1}
w^\dag w\leq\frac{1}{\gamma^2}z^\dag z+\delta.
\end{equation}
We define
\begin{equation}
\mathcal{W}_1=\left\{\begin{array}{c}H_2 :\exists \ w, z\ \textnormal{such that} \ (\ref{sector1})\ \textnormal {is satisfied}\\ \textnormal{and}\ (\ref{decomposition1})\ \textnormal{is satisfied}\ \forall \ V\in \mathcal{P}\end{array} \right\}.
\end{equation}

We then consider a set of quadratic perturbation Hamiltonians $\mathcal{W}_2$ that is in the form of
\begin{equation}\label{h2form1}
H_2=\frac{1}{2}\left[\begin{array}{c c} \zeta^\dag & \zeta^T\end{array}\right]\Delta\left[\begin{array}{l} \zeta \\ \zeta^\#\end{array}\right]
\end{equation}
where $\zeta=E_1a+E_2a^\#$ and $\Delta\in\mathbb{C}^{2m\times2m}$ is a Hermitian matrix of the form
\begin{equation}\label{delta1}
\Delta=\left[\begin{array}{cc}\Delta_1&\Delta_2\\ \Delta^\#_2&\Delta^\#_1\end{array}\right]
\end{equation}
with $\Delta_1=\Delta^\dag_1$ and $\Delta_2=\Delta^T_2$.
The matrix $\Delta$ is subject to the norm bound
\begin{equation}\label{delta_bound}
\| \Delta \|\leq\frac{2}{\gamma}
\end{equation}
where $\|.\|$ refers to the matrix induced norm.

We define
\begin{equation}
\mathcal{W}_2=\left\{\begin{array}{c}H_2 \textnormal { of the form (\ref{h2form1}) such that }\\
\textnormal{conditions (\ref{delta1}) and (\ref{commutation1}) are satisfied}\end{array} \right\}.
\end{equation}

Since the nominal system is linear, we use the relationship:
\begin{equation}\label{quadraticz}
z=\left[\begin{array}{l} \zeta \\ \zeta^\#\end{array}\right]=\left[\begin{array}{c c} E_1&E_2\\E^\#_1&E^\#_2\end{array}\right]\left[\begin{array}{l} a \\ a^\#\end{array}\right]=E\left[\begin{array}{l} a \\ a^\#\end{array}\right].
\end{equation}
Then
\begin{equation}
H_2=\frac{1}{2}\left[\begin{array}{c c} a^\dag & a^T\end{array}\right]E^\dag\Delta E\left[\begin{array}{l} a \\ a^\#\end{array}\right]
\end{equation}
In \cite{Ian2012}, it has been proven that for any set of self-adjoint operators $\mathcal{P}$,
\begin{equation}\label{w1relation}
\mathcal{W}_2\subset\mathcal{W}_1.
\end{equation}

\subsection{Non-quadratic Hamiltonian Perturbations}
For the set of non-negative self-adjoint operators $\mathcal{P}$ and given real parameters $\gamma>0$, $\delta_1\geq0$ and $\delta_2\geq0$, a particular set of perturbation Hamiltonians $\mathcal{W}_3$ is defined in terms of commutator decomposition
\begin {equation}\label{decomposition2}
[V,H_2]=[V, z]w^\ast_1-w_1[z^\ast,V]+\frac{1}{2}[z,[V,z]]w^\ast_2-\frac{1}{2}w_2[z,[V,z]]^\ast
\end{equation}
for $V\in \mathcal{P}$, where $w_1, w_2$ and $z$ are given scalar operators. Here, the notation $^\ast$ refers to the adjoint of an operator.
The set $\mathcal{W}_3$ is defined in terms of the sector bound condition
\begin{equation}\label{sector2}
w_1 w^\ast_1\leq\frac{1}{\gamma^2}zz^\ast+\delta_1
\end{equation}
and the condition
\begin{equation}\label{sector22}
w_2 w^\ast_2\leq\delta_2.
\end{equation}
We define
\begin{equation}
\mathcal{W}_3=\left\{\begin{array}{c}H_2 :\exists \ w_1, w_2, z\ \textnormal{such that} \  \textnormal{(\ref{sector2}) and\ (\ref{sector22})}\\ \textnormal{are satisfied and (\ref{decomposition2}) is satisfied}\ \forall \ V\in \mathcal{P}\end{array} \right\}.
\end{equation}

We consider a set of non-quadratic perturbation Hamiltonians $\mathcal{W}_4$. For the set of non-negative self-adjoint operators $\mathcal{P}$ and given real parameters $\gamma>0$, $\delta_1\geq0$, $\delta_2\geq0$, a set of non-quadratic perturbation Hamiltonians $\mathcal{W}_4$ is defined in terms of the following power series.
\begin{equation}\label{h2form2}
H_2=f(\zeta,\zeta^\ast)=\sum^\infty_{k=0}\sum^\infty_{l=0}S_{kl}\zeta^k(\zeta^\ast)^l=\sum^\infty_{k=0}\sum^\infty_{l=0}S_{kl}H_{kl}
\end{equation}
where $S_{kl}=S_{lk}^\ast$, $H_{kl}=\zeta^k(\zeta^\ast)^l$, and $\zeta$ is a scalar operator on the underlying Hilbert space.
Also
\begin{equation}
H_2^\ast=\sum^\infty_{k=0}\sum^\infty_{l=0}S_{kl}^\ast\zeta^l(\zeta^\ast)^k=\sum^\infty_{l=0}\sum^\infty_{k=0}S_{lk}\zeta^l(\zeta^\ast)^k=H_{2}.
\end{equation}
Hence, $H_{2}$ is self-adjoint operator.
We define
\begin{equation}
f'(\zeta,\zeta^\ast)=\sum^\infty_{k=1}\sum^\infty_{l=0}kS_{kl}\zeta^{k-1}(\zeta^\ast)^l,
\end{equation}
\begin{equation}
f''(\zeta,\zeta^\ast)=\sum^\infty_{k=1}\sum^\infty_{l=0}k(k-1)S_{kl}\zeta^{k-2}(\zeta^\ast)^l.
\end{equation}
We consider the sector bound condition
\begin{equation}\label{f'}
f'(\zeta,\zeta^\ast)^\ast f'(\zeta,\zeta^\ast)\leq\frac{1}{\gamma^2}\zeta\zeta^\ast+\delta_1,
\end{equation}
and the condition
\begin{equation}\label{f''}
f''(\zeta,\zeta^\ast)^\ast f''(\zeta,\zeta^\ast)\leq \delta_2.
\end{equation}
We define
\begin{equation}
\mathcal{W}_4=\left\{\begin{array}{c}H_2 \textnormal { of the form (\ref{h2form2}) such that (\ref{f'}) and }\\
\textnormal{ (\ref{f''})  are satisfied}\end{array} \right\}.
\end{equation}
From \cite{Ian2012},  we have the fact that for any set of self-adjoint operators $\mathcal{P}$,
\begin{equation}\label{w3relation}
\mathcal{W}_4\subset\mathcal{W}_3.
\end{equation}

In the nominal linear system, we define
\begin{equation}\label{nonquadraticz}
\begin{split}
z&=\zeta=E_1a+E_2a^\#\\
&=\left[\begin{array}{c c} E_1&E_2\end{array}\right]\left[\begin{array}{l} a \\ a^\#\end{array}\right]=\tilde{E}\left[\begin{array}{l} a \\ a^\#\end{array}\right].
\end{split}
\end{equation}
The following result has been proven in \cite{Ian2012}.
\begin{lemma}(See Lemma 5 of\cite{Ian2012})
Given any $V\in\mathcal{P}$,
\begin{equation}\label{mu}
\mu=[z,[V, z]]=-\tilde{E}\Sigma JPJ\tilde{E}^T \textnormal{is a constant,}
\end{equation}
where $\Sigma=\left[\begin{array}{cc} 0&I\\I&0\end{array}\right]$.
\end{lemma}

\section{PERFORMANCE ANALYSIS}\label{performance analysis}
In this section, we present several results on performance analysis for the two classes of quantum systems defined above.
We define the associated cost function for a quantum system as
\begin{equation}
J=\limsup \limits_{T\to \infty} \frac{1}{T}\int_0^T\langle \left[\begin{array}{c c} a^\dag & a^T\end{array}\right]R\left[\begin{array}{l} a \\ a^\#\end{array}\right]\rangle dt
\end{equation}
where $R>0$.
We denote that
\begin{equation}
W=\left[\begin{array}{c c} a^\dag & a^T\end{array}\right]R\left[\begin{array}{l} a \\ a^\#\end{array}\right].
\end{equation}
In order to prove the following  theorems on performance analysis, we require some algebraic identities.
\begin{lemma}\label{algebra}
(See Lemma 3 of\cite{Ian2012}) Consider $V\in\mathcal{P}, H_1$ is of the form (\ref{17}) and $L$ is of the form (\ref{19}). Then
\begin{equation}
[V, H_1]= \left[\begin{array}{l}a\\a^\#\end{array}\right]^\dag(PJM-MJP) \left[\begin{array}{l}a\\a^\#\end{array}\right],
\end{equation}
\begin{align}
\mathcal{L}(V)=&-\frac{1}{2} \left[\begin{array}{l}a\\a^\#\end{array}\right]^\dag(N^\dag JNJP+PJN^\dag JN)\left[\begin{array}{l}a\\a^\#\end{array}\right]\nonumber\\
&+\textnormal{Tr}(PJN^\dag \left[\begin{array}{cc} I&0\\0&0\end{array}\right]NJ),
\end{align}
\begin{equation}
[\left[\begin{array}{l}a\\a^\#\end{array}\right],\left[\begin{array}{l}a\\a^\#\end{array}\right]^\dag P\left[\begin{array}{l}a\\a^\#\end{array}\right]]=2JP\left[\begin{array}{l}a\\a^\#\end{array}\right].
\end{equation}
\end{lemma}

Now we present two theorems which can be used to analyse the performance of quantum systems subject to quadratic Hamiltonian perturbations and non-quadratic Hamiltonian perturbations, respectively.
\subsection{Quadratic Hamiltonian Perturbations}
\begin{theorem}\label{theorem1}
Consider an uncertain quantum system $(S, L, H)$, where $H=H_1+H_2$, $H_1$ is in the form of (\ref{17}),
$L$ is of the form (\ref{19}) and $H_2\in\mathcal{W}_2$. If $F=-iJM-\frac{1}{2}JN^\dag JN$ is Hurwitz, and
\begin{equation}\label{riccati}
\left[\begin{array}{cc}F^\dag P+PF+\frac{E^\dag E}{\gamma^2\tau^2}+R&2PJE^\dag\\ 2EJP&-I/\tau^2\end{array}\right]<0
\end{equation}
has a solution $P>0$ in the form of (\ref{18}) and $\tau>0$, then
\begin{align}
J&=\limsup \limits_{T\to \infty} \frac{1}{T}\int_0^T\langle W(t)\rangle dt\nonumber\\
&=\limsup \limits_{T\to \infty} \frac{1}{T}\int_0^T\langle \left[\begin{array}{c c} a^\dag & a^T\end{array}\right]R\left[\begin{array}{l} a \\ a^\#\end{array}\right]\rangle dt\leq\tilde{\lambda}+\frac{\delta}{\tau^2}
\end{align}
where
\begin{equation}
\tilde{\lambda}=\text{Tr}(PJN^\dag \left[\begin{array}{cc} I&0\\0&0\end{array}\right]NJ).
\end{equation}
\end{theorem}
In order to prove this theorem, we need the following two lemmas.
\begin{lemma}\label{lemmaw1}
Consider an open quantum system $( S, L, H)$ where $H=H_1+H_2$ and $H_2 \in \mathcal{W}_1$, and the set of non-negative self-adjoint operators $\mathcal{P}$. If there exists a $V \in \mathcal{P}$ and real constants $\tilde{\lambda}\geq0$, $\tau>0$ such that
\begin{equation}\label{lemmaw1equation}
-i[V,H_1]+\mathcal{L}(V)+\tau^2[V,z^\dag][z,V]+\frac{1}{\gamma^2\tau^2}z^\dag z+W\leq\tilde{\lambda},
\end{equation}
then
\begin{equation}
\limsup \limits_{T\rightarrow \infty} \frac{1}{T}\int_0^T\langle W(t)\rangle dt\leq\tilde{\lambda}+\frac{\delta}{\tau^2}, \forall t\geq0.
\end{equation}
\end{lemma}

\emph{Proof:}
Since $V\in \mathcal{P}$ and $H_2 \in \mathcal{W}_1$,
\begin{equation}\label{e9}
\mathcal{G}(V)=-i[V,H_1]+\mathcal{L}(V)-i[V, z^\dag]w+iw^\dag[z,V].
\end{equation}
Also,
\begin{equation}\label{e10}
\begin{split}
0&\leq(\tau[V,z^\dag]-\frac{i}{\tau}w^\dag)(\tau[V,z^\dag]-\frac{i}{\tau} w^\dag)^\dag\\
&=\tau^2[V,z^\dag][z,V]+i[V,z^\dag]w-iw^\dag[z,V]+\frac{w^\dag w}{\tau^2}.
\end{split}
\end{equation}
Substituting (\ref{e9}) into (\ref{e10}) and using the sector bound condition (\ref{sector1}), the following inequality is obtained:
\begin{equation}
\mathcal {G}(V)\leq-i[V,H_1]+\mathcal{L}(V)+\tau^2[V,z^\dag][z,V]+\frac{1}{\gamma^2\tau^2}z^\dag z+\frac{\delta}{\tau^2},
\end{equation}
Hence,
\begin{equation}
\mathcal {G}(V)+W\leq \tilde{\lambda}+\frac{\delta}{\tau^2}.
\end{equation}
Consequently, the conclusion in the lemma follows from Lemma \ref{1lemma}.
\hfill $\Box$
\begin{lemma}\label{quadratic_algebra}
For $V\in\mathcal{P}$ and $z$ defined in (\ref{quadraticz}),
\begin{equation}
[z,V]=2EJP\left[\begin{array}{l} a\\a^\# \end{array}\right],
\end{equation}

\begin{equation}
[V,z^\dag][z,V]=4\left[\begin{array}{l} a\\a^\# \end{array}\right]^\dag PJE^\dag EJP\left[\begin{array}{l} a\\a^\# \end{array}\right],
\end{equation}

\begin{equation}
z^\dag z=\left[\begin{array}{l} a\\a^\# \end{array}\right]^\dag E^\dag E\left[\begin{array}{l} a\\a^\# \end{array}\right].
\end{equation}
\end{lemma}
\emph{Proof:}
The proof follows from Lemma \ref{algebra}.
\hfill $\Box$\\
\emph{Proof of Theorem \ref{theorem1}}:
Using the Schur complement, the inequality ({\ref{riccati}) is equivalent to
\begin{equation}\label{40}
F^\dag P+PF+4\tau^2PJE^\dag EJP+\frac{E^\dag E}{\gamma^2\tau^2}+R<0.
\end{equation}
If the Riccati inequality (\ref{40}) has a solution $P>0$ of the form (\ref{18}) and $\tau>0$, according to Lemma \ref{algebra} and Lemma \ref{quadratic_algebra}, we have
\begin{equation}
\begin{split}
&-i[V,H_1]+\mathcal{L}(V)+\tau^2[V,z^\dag][z,V]+\frac{1}{\gamma^2\tau^2}z^\dag z+W=\\
&\left[\begin{array}{l} a\\a^\# \end{array}\right]^\dag\left(\begin{array}{l}F^\dag P+PF+4\tau^2PJE^\dag EJP\\+\frac{E^\dag E}{\gamma^2\tau^2}+R\end{array}\right)\left[\begin{array}{l} a\\a^\# \end{array}\right]\\
&\quad +\text{Tr}(PJN^\dag \left[\begin{array}{cc} I&0\\0&0\end{array}\right]NJ).
\end{split}
\end{equation}
Therefore, it follows from (\ref{riccati}) that condition (\ref{lemmaw1equation}) will be satisfied with
\begin{equation}
\tilde{\lambda}=\text{Tr}(PJN^\dag \left[\begin{array}{cc} I&0\\0&0\end{array}\right]NJ)\geq0.
\end{equation}
Then, according to the relationship (\ref{w1relation}) and Lemma \ref{lemmaw1}, we have
\begin{equation}
\begin{split}
&\limsup \limits_{T\to \infty} \frac{1}{T}\int_0^T\langle W(t)\rangle dt\\
&=\limsup \limits_{T\to \infty} \frac{1}{T}\int_0^T\langle \left[\begin{array}{c c} a^\dag & a^T\end{array}\right]R\left[\begin{array}{l} a \\ a^\#\end{array}\right]\rangle dt\leq\tilde{\lambda}+\frac{\delta}{\tau^2}.
\end{split}
\end{equation}
\hfill $\Box$
\subsection{Non-quadratic Hamiltonian Perturbations}
%
\begin{theorem}\label{theorem2}
Consider an uncertain quantum system $(S, L, H)$, where $H=H_1+H_2$, $H_1$ is in the form of (\ref{17}),
$L$ is of the form (\ref{19}) and $H_2\in\mathcal{W}_4$. If $F=-iJM-\frac{1}{2}JN^\dag JN$ is Hurwitz, and
\begin{equation}\label{nonriccati}
\left[\begin{array}{cc}F^\dag P+PF+\frac{\Sigma\tilde{E}^T\tilde{E}^\#\Sigma}{\gamma^2\tau^2}+R&2PJ\Sigma\tilde{E}^T\\ 2\tilde{E}^\#\Sigma JP&-I/\tau^2\end{array}\right]<0
\end{equation}
has a solution $P>0$ of the form (\ref{18}) and $\tau>0$, then
\begin{equation}
\begin{split}
J&=\limsup \limits_{T\to \infty} \frac{1}{T}\int_0^T\langle W(t)\rangle dt\\
&=\limsup \limits_{T\to \infty} \frac{1}{T}\int_0^T\langle \left[\begin{array}{c c} a^\dag & a^T\end{array}\right]R\left[\begin{array}{l} a \\ a^\#\end{array}\right]\rangle dt\\
&\leq\tilde{\lambda}+\frac{\delta_1}{\tau^2}+\mu\mu^\ast/4+\delta_2
\end{split}
\end{equation}
where
\begin{equation}
\tilde{\lambda}=\text{Tr}(PJN^\dag \left[\begin{array}{cc} I&0\\0&0\end{array}\right]NJ),
\end{equation}
and $\mu$ is defined as in (\ref{mu}).
\end{theorem}

In order to prove this theorem, we need the following two lemmas.
\begin{lemma}\label{lemmaw2}
Consider an open quantum system $( S, L, H)$ where $H=H_1+H_2$ and $H_2 \in \mathcal{W}_3$, and the set of non-negative self-adjoint operators $\mathcal{P}$. For any $V \in \mathcal{P}$, $\mu$ is a constant. If there exist real constants $\tilde{\lambda}\geq0$ and $\tau>0$ such that
\begin{equation}\label{lemmaw2equation}
-i[V,H_1]+\mathcal{L}(V)+\tau^2[V,z][z^\ast,V]+\frac{1}{\gamma^2\tau^2}zz^\ast+W\leq\tilde{\lambda},
\end{equation}
then
\begin{equation}
\limsup \limits_{T\rightarrow \infty} \frac{1}{T}\int_0^T\langle W(t)\rangle dt\leq\tilde{\lambda}+\frac{\delta_1}{\tau^2}+\mu\mu^\ast/4+\delta_2, \forall t\geq0.
\end{equation}
\end{lemma}
\emph{Proof:}
Since $V\in \mathcal{P}$ and $H_2 \in \mathcal{W}_3$ ,
\begin{align}\label{non9}
\mathcal{G}(V)=&-i[V,H_1]+\mathcal{L}(V)-i[V, z]w^\ast_1+iw_1[z^\ast,V]\nonumber\\
                            &-\frac{i}{2}\mu w^\ast_2+\frac{i}{2}w_2\mu^\ast.
\end{align}
Also,
\begin{equation}\label{non10}
\begin{split}
0&\leq(\tau[V,z]-\frac{i}{\tau}w_1)(\tau[V,z]-\frac{i}{\tau}w_1)^\ast\\
&=\tau^2[V,z][z^\ast,V]+i[V,z]w^\ast_1-iw_1[z^\ast,V]+\frac{1}{\tau^2}w_1w^\ast_1.
\end{split}
\end{equation}
Furthermore,
\begin{equation}\label{non11}
\begin{split}
0&\leq(\frac{1}{2}\mu-iw_2)(\frac{1}{2}\mu-iw_2)^\ast\\
&=\frac{1}{4}\mu\mu^\ast-\frac{i}{2}w_2\mu^\ast+\frac{i}{2}\mu w_2^\ast+w_2 w^\ast_2.
\end{split}
\end{equation}
Substituting (\ref{non10}) and (\ref{non11}) into (\ref{non9}), the following inequality is obtained.
\begin{align}
\mathcal {G}(V)\leq&-i[V,H_1]+\mathcal{L}(V)+\tau^2[V,z][z^\ast,V]+\frac{1}{\gamma^2\tau^2}zz^\ast\nonumber\\
&+\frac{\delta_1}{\tau^2}+\frac{1}{4}\mu\mu^\ast+\delta_2.
\end{align}
Hence, using (\ref{lemmaw2equation}), it follows that
\begin{equation}
\mathcal {G}(V)+W\leq\tilde{\lambda}+\frac{\delta_1}{\tau^2}+\mu\mu^\ast/4+\delta_2.
\end{equation}
Consequently, the conclusion in the lemma follows from Lemma \ref{1lemma}.
\hfill $\Box$

\begin{lemma}\label{nonquadratic_algebra}
For $V\in\mathcal{P}$ and $z$ defined in (\ref{nonquadraticz}),
\begin{equation}
[z^\ast,V]=2\tilde{E}^\#\Sigma JP\left[\begin{array}{l} a\\a^\# \end{array}\right],
\end{equation}

\begin{equation}
[V,z][z^\ast,V]=4\left[\begin{array}{l} a\\a^\# \end{array}\right]^\dag PJ\Sigma\tilde{E}^{T}\tilde{E}^\#\Sigma JP\left[\begin{array}{l} a\\a^\# \end{array}\right],
\end{equation}

\begin{equation}
zz^\ast=\left[\begin{array}{l} a\\a^\# \end{array}\right]^\dag \Sigma\tilde{E}^{T} \tilde{E}^\#\Sigma\left[\begin{array}{l} a\\a^\# \end{array}\right].
\end{equation}

\end{lemma}
\emph{Proof:}
The proof follows from Lemma \ref{algebra}.
\hfill $\Box$\\
\emph{Proof of Theorem \ref{theorem2}:}
Using the Schur complement, the inequality (\ref{nonriccati}) is equivalent to
\begin{equation}\label{taoinequality}
F^\dag P+PF+4\tau^2PJ\Sigma\tilde{E}^{T}\tilde{E}^\#\Sigma JP+\Sigma\tilde{E}^{T}\tilde{E}^\#\Sigma/(\gamma^2\tau^2)+R<0.
\end{equation}
If the Riccati inequality (\ref{taoinequality}) has a solution $P>0$ in the form of (\ref{18}) and $\tau>0$, according to Lemma \ref{algebra} and Lemma \ref{nonquadratic_algebra}, we have
\begin{align}
&-i[V,H_1]+\mathcal{L}(V)+\tau^2[V,z][z^\ast,V]+\frac{1}{\gamma^2\tau^2}zz^\ast+W=\nonumber\\
&\left[\begin{array}{l} a\\a^\# \end{array}\right]^\dag\left(\begin{array}{c}F^\dag P+PF+4\tau^2PJ\Sigma\tilde{E}^{T}\tilde{E}^\#\Sigma JP\\+\Sigma\tilde{E}^{T}\tilde{E}^\#\Sigma/(\gamma^2\tau^2)+R\end{array}\right)\left[\begin{array}{l} a\\a^\# \end{array}\right]\nonumber\\
&\quad +\text{Tr}(PJN^\dag \left[\begin{array}{cc} I&0\\0&0\end{array}\right]NJ).
\end{align}
Therefore, it follows from (\ref{nonriccati}) that condition (\ref{lemmaw2equation}) will be satisfied with
\begin{equation}
\tilde{\lambda}=\text{Tr}(PJN^\dag \left[\begin{array}{cc} I&0\\0&0\end{array}\right]NJ)\geq0.
\end{equation}
According to the relationship (\ref{w3relation}) and Lemma \ref{lemmaw2}, we have
\begin{equation}
\begin{split}
&\limsup \limits_{T\to \infty} \frac{1}{T}\int_0^T\langle W(t)\rangle dt\\
&=\limsup \limits_{T\to \infty} \frac{1}{T}\int_0^T\langle \left[\begin{array}{c c} a^\dag & a^T\end{array}\right]R\left[\begin{array}{l} a \\ a^\#\end{array}\right]\rangle dt\\
&\leq\tilde{\lambda}+\frac{\delta_1}{\tau^2}+\mu\mu^\ast/4+\delta_2.
\end{split}
\end{equation}
\hfill $\Box$

\section{COHERENT GUARANTEED COST CONTROLLER DESIGN}\label{optimal quantum controllers design}
In some applications, it is desirable to design a quantum control system which is not only stable but also guarantees an adequate level of performance.
In this section, we design a coherent guaranteed cost controller for quantum systems subject to quadratic or non-quadratic Hamiltonian perturbations. The coherent controller is realized by adding a term $H_3$ to the nominal system Hamiltonian.
Assume that the controller Hamiltonian $H_3$ is of the form
\begin{equation}\label{H3}
H_3=\frac{1}{2}\left[\begin{array}{c c} a^\dag & a^T\end{array}\right]K\left[\begin{array}{l} a \\ a^\#\end{array}\right]
\end{equation}
where $K\in\mathbb{C}^{2n\times2n}$ is a Hermitian matrix of the form
\begin{equation}
K=\left[\begin{array}{cc}K_1&K_2\\K_2^\#&K_1^\#\end{array}\right]
\end{equation}
and $K_1=K^\dag_1$, $K_2=K^T_2$.
Associated with this system is the cost function $J$
\begin{equation}
J=\limsup \limits_{T\to \infty} \frac{1}{T}\int_0^\infty (\left[\begin{array}{l} a \\ a^\#\end{array}\right]^\dag (R+ \rho K^2)\left[\begin{array}{l} a \\ a^\#\end{array}\right])dt
\end{equation}
where $\rho\in(0,\infty)$ is a weighting factor.
We let
\begin{equation}
W=\left[\begin{array}{l} a \\ a^\#\end{array}\right]^\dag (R+\rho K^2)\left[\begin{array}{l} a \\ a^\#\end{array}\right].
\end{equation}

The following sections present our main results on coherent guaranteed cost controller design for quantum systems subject to quadratic and non-quadratic Hamiltonian perturbations, respectively.
\subsection{Quadratic Hamiltonian Perturbations}
\begin{theorem}\label{theoremcontrol}
Consider an uncertain quantum system $(S, L, H)$, where $H=H_1+H_2+H_3$, $H_1$ is in the form of (\ref{17}), $L$ is of the form (\ref{19}), $H_2\in\mathcal{W}_2$ and the controller Hamiltonian $H_3$ is in the form of (\ref{H3}). 
With $Q=P^{-1}$, $Y=KQ$ and $F=-iJM-\frac{1}{2}JN^\dag JN$, if there exist a matrix $Q =q*I$ ($q$ is a constant and $I$ is identity matrix), a Hermitian matrix $Y$ and a constant $\tau>0$, such that
\begin{equation}\label{lmi}
\left[\begin{array}{cccc} A+4\tau^2JE^\dag EJ&Y& qR^{\frac{1}{2}}&qE^\dag
\\Y&-I/\rho &0&0\\qR^{\frac{1}{2}}&0&-I&0\\qE&0&0&-\gamma^2\tau^2I\end{array}\right]<0
\end{equation}
where $A=qF^\dag+Fq+iYJ-iJY$,
then the associated cost function satisfies the bound
\begin{equation}\label{optimal_controller}
\begin{split}
&\limsup \limits_{T\to \infty} \frac{1}{T}\int_0^T\langle W(t)\rangle dt\\
&=\limsup \limits_{T\to \infty} \frac{1}{T}\int_0^\infty (\left[\begin{array}{l} a \\ a^\#\end{array}\right]^\dag (R+ \rho K^2)\left[\begin{array}{l} a \\ a^\#\end{array}\right])dt\leq\tilde{\lambda}+\frac{\delta}{\tau^2}
\end{split}
\end{equation}
where
\begin{equation}
\tilde{\lambda}=\text{Tr}(PJN^\dag \left[\begin{array}{cc} I&0\\0&0\end{array}\right]NJ).
\end{equation}
\end{theorem}
\emph{Proof:}
Suppose the conditions of the theorem are satisfied. Using the Schur complement, (\ref{lmi}) is equivalent to
\begin{equation}\label{eqf2}
\left[\begin{array}{ccc} A+4\tau^2JE^\dag EJ+\rho YY& qR^{\frac{1}{2}}&qE^\dag
\\qR^{\frac{1}{2}}&-I&0\\qE&0&-\gamma^2\tau^2I\end{array}\right]<0.
\end{equation}
Applying the Schur complement again, it is clear that (\ref{eqf2}) is equivalent to
\begin{equation}\label{eqf11}
\left[\begin{array}{cc} A+4\tau^2JE^\dag EJ+\rho YY+q^2R&qE^\dag\\ qE&-\gamma^2\tau^2I \end{array}\right]<0
\end{equation}
and (\ref{eqf11}) is equivalent to
\begin{equation}\label{eqf1}
\begin{split}
&qF^\dag+Fq+iYJ-iJY+4\tau^2JE^\dag EJ\\
&+\rho YY+q^2(\frac{E^\dag E}{\gamma^2\tau^2}+R)<0.
\end{split}
\end{equation}
Substituting $Y=Kq=qK^\dag$ into (\ref{eqf1}), we obtain
\begin{equation}
\begin{split}
&q(F-iJK)^\dag +(F-iJK)q+4\tau^2JE^\dag EJ\\
&+q^2(\frac{E^\dag E}{\gamma^2\tau^2}+R+\rho K^2)<0.
\end{split}
\end{equation}
Since $P=Q^{-1}$, premultiplying and postmultiplying this inequality by the matrix $P$, we have
\begin{equation}
\begin{split}
&(F-iJK)^\dag P+P(F-iJK)+4\tau^2PJE^\dag EJP\\
&+\frac{E^\dag E}{\gamma^2\tau^2}+R+\rho K^2<0.
\end{split}
\end{equation}
We know $E^\dag E\geq0, R>0$ and $K^2>0$. Hence\\
\begin{equation}
\begin{split}
4\tau^2PJE^\dag EJP+\frac{E^\dag E}{\gamma^2\tau^2}+R+\rho K^2>0,\\
(F-iJK)^\dag P+P(F-iJK)<0.
\end{split}
\end{equation}
Therefore, we have the following fact
\begin{equation}
\tilde{F}=-iJ(M+K)-\frac{1}{2}JN^\dag JN \textnormal{ is Hurwitz}.
\end{equation}
We also know that
\begin{align}
&-i[V,H_1+H_3]+\mathcal{L}(V)+\tau^2[V,z^\dag][z,V]+\frac{1}{\gamma^2\tau^2}z^\dag z+W\nonumber\\
&=\left[\begin{array}{l} a\\a^\# \end{array}\right]^\dag\left(\begin{array}{l}(F-iJK)^\dag P+P(F-iJK)\nonumber\\
+4\tau^2PJE^\dag EJP+E^\dag E/(\gamma^2\tau^2)\\+R+\rho K^2
\end{array}\right)\left[\begin{array}{l} a\\a^\# \end{array}\right]\nonumber\\
&\quad +\text{Tr}(PJN^\dag \left[\begin{array}{cc} I&0\\0&0\end{array}\right]NJ).
\end{align}
According to the relationship (\ref{w1relation}) and Lemma \ref{lemmaw1}, we have
\begin{equation}
\begin{split}
&\limsup \limits_{T\to \infty} \frac{1}{T}\int_0^T\langle W(t)\rangle dt\\
&=\limsup \limits_{T\to \infty} \frac{1}{T}\int_0^\infty (\left[\begin{array}{l} a \\ a^\#\end{array}\right]^\dag (R+ \rho K^2)\left[\begin{array}{l} a \\ a^\#\end{array}\right])dt\\
&\leq\tilde{\lambda}+\frac{\delta}{\tau^2}
\end{split}
\end{equation}
where
\begin{equation}
\tilde{\lambda}=\text{Tr}(PJN^\dag \left[\begin{array}{cc} I&0\\0&0\end{array}\right]NJ).
\end{equation}
\hfill $\Box$
\begin{remark}\label{remark1}
In order to design a coherent controller which minimizes the cost bound (\ref{optimal_controller}) in the above theorem, we need to formulate an inequality
\begin{equation}\label{quadraticxi}
\text{Tr}(PJN^\dag \left[\begin{array}{cc} I&0\\0&0\end{array}\right]NJ)+\frac{\delta}{\tau^2}\leq\xi.
\end{equation}
We know that $P=Q^{-1}=q^{-1}I$ and apply Schur complement to inequality (\ref{quadraticxi}), so that we have
\begin{equation}\label{qudraticxi1}
\left[\begin{array}{cc} -\xi+\frac{\delta}{\tau^2}&B^\frac{1}{2}\\ B^\frac{1}{2}&-q\end{array}\right]\leq0
\end{equation}
where $B=\text{Tr}(JN^\dag \left[\begin{array}{cc} I&0\\0&0\end{array}\right]NJ).$
Applying Schur Complement again, it is clear that (\ref{qudraticxi1}) is equivalent to
\begin{equation}\label{qudraticxi2}
\left[\begin{array}{ccc} -\xi&\delta^\frac{1}{2}&B^\frac{1}{2}\\ \delta^\frac{1}{2}&-\tau^2&0\\B^\frac{1}{2}&0&-q\end{array}\right]\leq0.
\end{equation}
Hence, we minimize $\xi$ subject to  (\ref{qudraticxi2}) and (\ref{lmi}) in Theorem \ref{theoremcontrol}. This is a standard LMI problem.
\end{remark}
\subsection{Non-quadratic Hamiltonian Perturbations}
\begin{theorem}\label{nonquadratichamiltoniancontroller}
Consider an uncertain quantum system $(S, L, H)$, where $H=H_1+H_2+H_3$, $H_1$ is in the form of (\ref{17}), $L$ is of the form (\ref{19}), $H_2\in\mathcal{W}_4$ and the controller Hamiltonian $H_3$ is in the form of (\ref{H3}). 
With $Q=P^{-1}$, $Y=KQ$ and $F=-iJM-\frac{1}{2}JN^\dag JN$, if there exist a matrix $Q=q*I$, a Hermitian matrix $Y$ and a constant $\tau>0$, such that
\begin{equation}\label{nlmi}
\left[\begin{array}{cccc} A+4\tau^2J\Sigma\tilde{E}^T\tilde{E}^\#\Sigma J&Y& qR^{\frac{1}{2}}&q\Sigma\tilde{E}^T
\\Y&-I/\rho &0&0\\qR^{\frac{1}{2}}&0&-I&0\\q\tilde{E}^\#\Sigma&0&0&-\gamma^2\tau^2I\end{array}\right]<0
\end{equation}
where $A=qF^\dag+Fq+iYJ-iJY$,
then the associated cost function satisfies the bound
\begin{equation}\label{noptimal_controller}
\limsup \limits_{T\to \infty} \frac{1}{T}\int_0^T\langle W(t)\rangle dt\leq\tilde{\lambda}+\frac{\delta_1}{\tau^2}+\mu\mu^\ast/4+\delta_2
\end{equation}
where
\begin{equation}
\tilde{\lambda}=\text{Tr}(PJN^\dag \left[\begin{array}{cc} I&0\\0&0\end{array}\right]NJ),
\end{equation}
\begin{equation}
\mu\mu^\ast/4=\frac{1}{4}\tilde{E}\Sigma JPJ\tilde{E}^T \tilde{E}^\#JPJ\Sigma\tilde{E}^\dag.
\end{equation}
\end{theorem}

\emph{Proof:}
Suppose the conditions of the theorem are satisfied. By using the same method as in the proof of Theorem \ref{theoremcontrol}, we have
\begin{equation}\label{95}
\begin{split}
&(F-iJK)^\dag P+P(F-iJK)+4\tau^2PJ\Sigma\tilde{E}^{T}\tilde{E}^\#\Sigma JP\\
&+\Sigma\tilde{E}^{T}\tilde{E}^\#\Sigma/(\gamma^2\tau^2)+R+\rho K^2<0.
\end{split}
\end{equation}
Using a similar method to Theorem \ref{theoremcontrol}, the following inequality is obtained.
\begin{equation}
(F-iJK)^\dag P+P(F-iJK)<0.
\end{equation}
Therefore, we have the following result:
\begin{equation}
\tilde{F}=-iJ(M+K)-\frac{1}{2}JN^\dag JN \textnormal{ is Hurwitz}.
\end{equation}
We also know that
\begin{align}
&-i[V,H_1+H_3]+\mathcal{L}(V)+\tau^2[V,z^\dag][z^\ast,V]+\frac{1}{\gamma^2\tau^2}zz^\ast+W\nonumber\\
&=\left[\begin{array}{l} a\\a^\# \end{array}\right]^\dag\left(\begin{array}{l}(F-iJK)^\dag P+P(F-iJK)\nonumber\\
+4\tau^2PJ\Sigma\tilde{E}^{T}\tilde{E}^\#\Sigma JP\\+\Sigma\tilde{E}^{T}\tilde{E}^\#\Sigma/(\gamma^2\tau^2)+R+\rho K^2
\end{array}\right)\left[\begin{array}{l} a\\a^\# \end{array}\right]\nonumber\\
&\quad +\text{Tr}(PJN^\dag \left[\begin{array}{cc} I&0\\0&0\end{array}\right]NJ).
\end{align}
It follows from (\ref{95}) that condition (\ref{lemmaw2equation}) will be satisfied with
\begin{equation}
\tilde{\lambda}=\text{Tr}(PJN^\dag \left[\begin{array}{cc} I&0\\0&0\end{array}\right]NJ).
\end{equation}
Therefore, the following result follows from the relationship (\ref{w3relation}) and Lemma \ref{lemmaw2}:
\begin{equation}
\limsup \limits_{T\to \infty} \frac{1}{T}\int_0^T\langle W(t)\rangle dt\leq\tilde{\lambda}+\frac{\delta_1}{\tau^2}+\mu\mu^\ast/4+\delta_2.
\end{equation}
\hfill $\Box$

%
\section{ILLUSTRATIVE EXAMPLE}\label{illustrative example}
Consider an example of an open quantum system with
\begin{equation}
H_1=\frac{1}{4}i((a^\dag)^2-a^2),
H_2=\frac{1}{4}i((a^\dag)^2-a^2),
S=I,
L=\sqrt{\kappa}a.
\end{equation}
The corresponding parameters considered in Theorems \ref{theorem1} and \ref{theoremcontrol} are as follows:
\begin{equation}
\begin{split}
&M=\left[\begin{array}{cc} 0&\frac{1}{2}i \\ -\frac{1}{2}i&0\end{array}\right],
N=\left[\begin{array}{cc} \sqrt{\kappa}&0\\ 0&\sqrt{\kappa}\end{array}\right],\\
&F=\left[\begin{array}{cc} -\frac{\kappa}{2}&0.5\\ 0.5&-\frac{\kappa}{2}\end{array}\right],
E=I, \delta=1
\end{split}
\end{equation}
and
\begin{equation}
\Delta\Delta=\left[\begin{array}{cc} 0&\frac{1}{2}i \\ -\frac{1}{2}i&0\end{array}\right]\left[\begin{array}{cc} 0&\frac{1}{2}i \\ -\frac{1}{2}i&0\end{array}\right]=\left[\begin{array}{cc} \frac{1}{4} &0\\ 0&\frac{1}{4}\end{array}\right].
\end{equation}
Here, $\gamma=1$ is chosen to satisfy (\ref{delta_bound}) and $H_2\in \mathcal{W}_2$. The dash line in Figure 1 is the cost bound for the linear quantum system  considered in Theorem \ref{theorem1} as a function of parameter $\kappa$.

Next, we add the guaranteed cost controller considered in Theorem \ref{theoremcontrol} to the linear quantum system. The solid line in Figure 1 shows the performance with the coherent controller. Compared to the performance of the quantum system without a controller, our coherent controller can guarantee the system is stable for a larger range of value $\kappa$
and leads to a closed loop system having better performance.
\begin{figure}[htb]
       \centering
        \includegraphics[width=0.5  \textwidth]{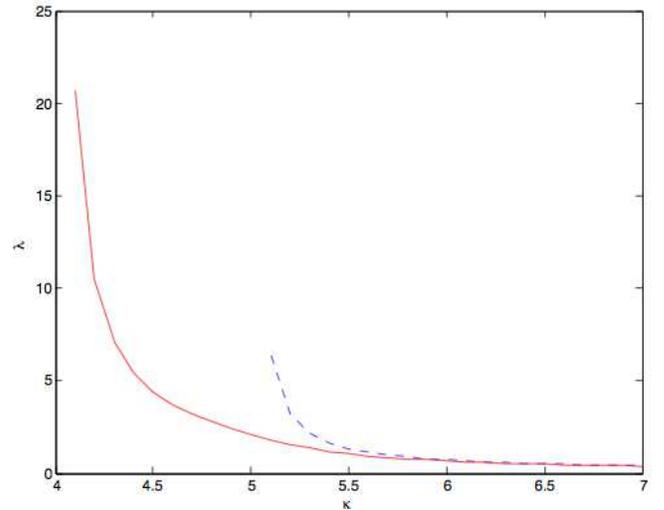}
        \caption{Guaranteed cost bounds for uncertain quantum systems with a controller (solid line) and without a controller (dash line)}
        \label{fig1}
\end{figure}



\section{CONCLUSION}\label{conclusion}
This paper has evaluated the performance of uncertain quantum systems with either quadratic or non-quadratic perturbations in the system Hamiltonian. We designed a coherent guaranteed cost controller for the class of uncertain quantum systems to obtain improved performance bounds.

\end{document}